\documentclass[twocolumn,secnumarabic,amssymb, amsmath, nobibnotes, aps, prl, nofootinbib]{revtex4-2}
\usepackage{graphics}      
\usepackage{graphicx}      
\usepackage{longtable}     
\usepackage{bm}            
\usepackage{natbib}
\usepackage{physics}
\usepackage{xcolor}
\usepackage[caption=false]{subfig}
\usepackage{dcolumn}
\usepackage{hyperref}
\usepackage{amsmath}
\usepackage{amssymb}
\usepackage{slashed}
\usepackage{mathtools}
\usepackage[thinc]{esdiff}
\usepackage{multirow}
\usepackage{float}
\usepackage{tabularx}

\begin{document}
	
	\title{Self-energy correction to energy levels of highly charged ions in a\\path integral formalism}%
	
	\author{S. Banerjee}%
	\email{banerjee@mpi-hd.mpg.de}
	\affiliation{Max Planck Institute for Nuclear Physics, Heidelberg, Germany}
	\author{Z. Harman}
	\email{harman@mpi-hd.mpg.de}
	\affiliation{Max Planck Institute for Nuclear Physics, Heidelberg, Germany}
	\date{\today}%

	\begin{abstract}
		Self-energy corrections to the energy levels of bound electrons are calculated in the framework of path integrals. We arrive at the full
		fermion propagator, using methods of functional integrals, in the form of Schwinger-Dyson equation (SDE). From the full fermion SDE, the
		self-energy corrected propagator is identified and the energy shift is obtained from the poles of the spectral function. The numerical
		calculations are performed using complex contour integrals and the B-spline representation of basis functions. We identify ions with
		Lamb shifts observable via modern mass spectrometric methods.
	\end{abstract}

	\maketitle

\section{Introduction}

Radiative corrections to energy levels have been at the focal point of theoretical and experimental studies with atoms, and, more recently, highly charged ions (HCI).
Experimental advances in the production of HCI and the measurement of their properties with unprecedented accuracy
(see e.g. Refs.~\cite{King2022,Sailer_2022,Micke2020,Rischka2020,Sch_ssler_2020,Kubicek2014,Sturm2011,Gillaspy2010,Nakamura2008,Gumberidze2005,Brandau2002,
Gillaspy1996,PhysRevLett.85.3109,Marmar1986,Deslattes1984, Beiersdorfer1989,Decaux1997,Tavernier1985}) call for versatile theoretical frameworks for the study of such systems.
The calculations pertaining to radiative corrections in quantum electrodynamics (QED), most importantly, the self-energy (SE) effect, have been rigorously studied
and developed by a multitude of authors
\cite{Uehling1935,Wichmann1956,Desiderio1971,Mohr1974,Mohr1975,Gyulassy1975,Snyderman1991,Indelicato1992,PhysRevA.46.3762,Persson1993,Pachucki1993,Pachucki1994,
Mohr1998,Jentschura1999,PhysRevA.13.1283,Yerokhin1999,Plunien1999,Shabaev_2000,Yerokhin2004,Yerokhin2021,RevModPhys.70.55}.
Historically, one of the first such studies was performed by Gell-Mann, Low and Sucher using the adiabatic $S$-matrix method~\cite{PhysRev.84.350,PhysRev.107.1448}. Later, a plethora of
theoretical formalisms based on Green's functions were developed~\cite{Shabaev1991,Shabaev1990,Shabaev_2002,braun1984relativistic,FELDMAN198720,Yennie:1988ju,FELDMAN1988231,Adkins:1988jc}. 

In this Letter, we arrive at the SE corrected fermion propagator in the Furry picture using the method of functional integrals conceived by Feynman~\cite{PhysRev.76.769}
and developed further by several authors, both in non-relativistic and relativistic quantum mechanics
\cite{Feynman2010,BHAGWAT1989417,Barut_1987,CAMBLONG_2004,Duru1979,INOMATA1982387,Ferraro:1992eb,Inomata1984,Peak1969}. The SE correction to energy levels
is evaluated for the first time in the path integral framework. We derive the dressed bound-fermion propagator using the Schwinger-Dyson equations (SDE).
The derivation of the energy shift induced by the interaction of the bound fermion with its own electromagnetic field is performed by separating the expression into the so-called
zero-, one-, and many-potential terms~\cite{Snyderman1991}, each of which are defined using perturbative path integrals. The finite contributions due to these individual
terms are calculated numerically using complex contour integrals, extensively worked out in Refs.~\cite{Snyderman1991,PhysRevA.46.3762,Yerokhin1999}. Computations are performed
by the known B-spline representation of bound-electron basis states with existing numerical methods~\cite{Cakir:2020uuy}.

The introduction of functional methods in atomic physics is also motivated by the prospects of including non-electromagnetic interactions
into precision theory. For example, hadronic vacuum polarization effects have been calculated by means of a quantum chromodynamic (QCD)
Schwinger-Dyson approach~\cite{Goecke2011,Bashir2012}. The improvement of experimental accuracy may necessitate in future the inclusion of such QCD corrections in atomic
spectra~\cite{Friar1999,Karshenboim2021,Breidenbach2022}. Prospects of new physics searches with low-energy atomic precision experiments
(see e.g.~\cite{Jaeckel2010,Flambaum2018,Berengut2018,Frugiuele2017,Counts2020,Sailer_2022}) also suggest to employ a versatile formalism enabling the description of various
types of exchange bosons.

\section{Schwinger-Dyson equation for the bound-fermion propagator}

We begin by deriving the complete expression for the dressed bound-fermion propagator.
This is accomplished by defining the SDE for the fermionic propagator using path integrals, in analogy to Ref.~\cite{PhysRevD.83.045007}. The QED Lagrangian
is given as
\begin{eqnarray}
	\mathcal{L}_{\text{QED}}(x)=&& -\frac{1}{4}F^{\mu\nu}F_{\mu\nu}-\frac{1}{2\xi}(\partial^\mu A_\mu)^2\label{Eq:1}\\
	&&+\bar{\psi}(x)(i\slashed{D}-m)\psi(x)-\bar{\psi}(x)e\gamma^\mu A_{\mu}(x)\psi(x)\nonumber\,,
\end{eqnarray}
where $D_\mu(x)=\partial_\mu(x)+ie\mathcal{A}_\mu(x)$,
$\mathcal{A}_\mu$ is the field of the nucleus; $A_\mu$ is the gauge-field operator for the photon field, $\psi(x)$ is the field of the electron, $m$ is its bare mass,
$e$ is the elementary charge, and $ F^{\mu\nu}=\partial^\mu A^{\nu}-\partial^{\nu} A^{\mu} $ is the electromagnetic field operator or the curvature of the field. The $\gamma$ are the
usual Dirac matrices and $ \mu,\nu \in \{0,1,2,3\} $ represent the Lorentz indices. We apply the background field method~\cite{PhysRev.162.1195, Abbott:1980hw, Abbott:1981ke}, where we consider the
external field to be the classical background field of the nuclear charge that is gauge invariant, and the photon gauge field is treated as a fluctuation whose gauge has been
fixed as seen in the second term of Eq.~(\ref{Eq:1}), where $ \xi $ is the gauge-fixing coefficient. We do not need to concern
ourselves with fixing the gauge for the external field since the effective action in the presence or absence of a classical background field is equivalent, and hence the choice
of gauge plays no role. We can also safely ignore this gauge-fixing term since we are interested in deriving the electron propagator in the presence of an external field,
and it does not contribute to the effective action of the theory.

The generating functional is constructed using the above Lagrangian, and Grassmann-valued sources $\eta$ and $\bar{\eta}$ of the fermion fields $\bar{\psi}$ and $\psi$,
respectively, and the source $J_\mu$ for the gauge field:
\begin{eqnarray}
	Z[\eta,\bar{\eta},J_\mu]&&=\int\mathcal{D}\bar{\psi}\mathcal{D}\psi\mathcal{D}A \exp \biggl\{i\int d^4 x[\mathcal{L}_{\text{QED}}\label{Eq:2}\\
	&&+J_\mu(x) A^\mu(x)+\bar{\psi}(x)\eta(x)+\bar{\eta}(x)\psi(x)]\biggr\}\,,\nonumber
\end{eqnarray}
where $ \mathcal{D} $ represents the integral measure over all field configurations.
To arrive at the SDE, we consider that the functional integral of a total derivative is zero, i.e.,
\begin{eqnarray}
	\int \mathcal{D}[\phi]\fdv{\phi}=0\,,\nonumber
\end{eqnarray}
where $ \fdv{\phi} $ is the functional derivative w.r.t. $\phi$, which represents any arbitrary field variable.
For the electron propagator, the derivative is taken with respect to the
fermion field $\bar{\psi}(x)$,
\begin{eqnarray}
	&&\int\mathcal{D}[\bar{\psi}\psi A]\frac{\delta}{\delta \bar{\psi}(x)}\exp\{i[\mathcal{S}(\bar{\psi},\psi, A)\label{Eq:3}\\
	&&\quad+\int \dd[4]{x}(J_\mu(x) A^\mu(x)+\bar{\psi}(x)\eta(x)+\bar{\eta}(x)\psi(x))]\}=0\,,\nonumber
\end{eqnarray}
with $ \mathcal{S}=\int \dd[4]{x} \mathcal{L}_{\text{QED}}$ being the action.
Eq.~(\ref{Eq:3}) can be written in terms of a differential equation in the generating functional $Z$ 
\begin{eqnarray}
	\Biggl[\frac{\delta \mathcal{S}}{\delta \bar{\psi}(x)}&&\left(-i\frac{\delta}{\delta J_\mu }, i\frac{\delta}{\delta \eta}, -i\frac{\delta}{\delta \bar{\eta}}\right)\label{Eq:4}\\
	&&\qquad\qquad\qquad+\eta(x)\Biggr]Z[\eta,\bar{\eta},J_\mu]=0\,.\nonumber
\end{eqnarray}
We take the functional derivative of the action $\mathcal{S}$ by implementing the G\^ateux derivative method~\cite{GT,article}, and arrive at the differential equation
\begin{eqnarray}
	\Biggl[&&\Biggl(i\slashed{\partial}-m-e\gamma^\mu \mathcal{A}_\mu-e\gamma^\mu(-i)\fdv{J^{\mu}(x)}\Biggr)  \label{Eq:5}\\
	&&\times (-i)\fdv{\bar{\eta}(x)}+\eta(x)\Biggr]Z[\eta, \bar{\eta}, J_\mu]=0\,.\nonumber
\end{eqnarray}
To obtain the two-point Green's function of the electron, we perform another derivation with respect to the source field $\eta(y)$
and represent it in terms of the functional for the connected Green's functions $W$, ($ Z=e^W $)
\begin{eqnarray}
	&&e^{W[\eta, \bar{\eta}, J_\mu]}\biggl[\delta(x-y)-\biggl(i\slashed{\partial}-m-e\gamma^\mu \mathcal{A}_\mu\label{Eq:6}\\
	&&-ie\gamma^\mu\frac{(-i)\delta W}{\delta J^\mu(x)}-ie\gamma^\mu(-i)\fdv{J^{\mu}(x)}\biggr)S(x,y)\biggr]=0\,.\nonumber
\end{eqnarray}
Here, $ S(x,y) $ is the bound-fermion propagator. We rewrite Eq.~(\ref{Eq:6}) in terms of classical fields, in analogy to~\cite{PhysRevD.83.045007}
\begin{eqnarray}
	&&\delta(x-y)-\biggl(i\slashed{\partial}-m-e\gamma^\mu \mathcal{A}_\mu\label{Eq:7}\\
	&&-ie\gamma^\mu A_\mu-ie\gamma^\mu(-i)\fdv{J^{\mu}(x)}\biggr)S(x,y)=0\,.\nonumber
\end{eqnarray}
We obtain the equation
\begin{eqnarray}
	&&(-i)\frac{\delta S(x,y)}{\delta J^{\mu}(x)}\label{Eq:8}\\
	&&\qquad=-i\int \dd[4]{z}\frac{\delta A_\nu(z)}{\delta J^\mu(x)}\fdv{A_\nu(z)}\left(\frac{\delta^2 \Gamma}{\delta \psi(x)\delta \bar{\psi}(y)}\right)^{-1}\nonumber\,.
\end{eqnarray}
Using the expressions for the complete bound-electron propagator, photon propagator and the electron-photon vertex
function $\Gamma^{\mu}$, one obtains
\begin{eqnarray}
	&&(-i)\frac{\delta S(x,y)}{\delta J^{\mu}(x)}\label{Eq:9}\\
	&&=-e\int \dd[4]{z}\dd[4]{u}\dd[4]{w}D_{\mu\nu}(x-z)S(x,w)\Gamma^\nu(w,u;z)S(u,y)\,.\nonumber
\end{eqnarray}
Using Eqs.~(\ref{Eq:8}) and (\ref{Eq:9}), setting the external source fields equal to zero, and considering that the nuclear field is a static scalar field,
Eq.~(\ref{Eq:7}) reduces to
\begin{eqnarray}
	\delta(x-y)&=&(i\slashed{\partial}-m-e\gamma^0 \mathcal{A}_{0})S(x,y)\label{Eq:10}\\
	&&+\int \dd[4]{u} \Sigma(x,u)S(u,y)\,,\nonumber
\end{eqnarray}
where 
\begin{equation}
	\Sigma(x-y)=-ie^2\gamma^\mu\int \dd[4]{z}\dd[4]{w}D_{\mu\nu}(z-x)S(x,w)\Gamma^\nu(w,y;z)\,.
	\label{Eq:11}
\end{equation}

\begin{figure}
	\centering
	\includegraphics[width=0.45\textwidth]{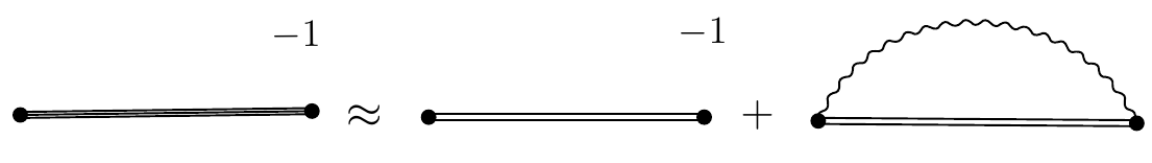}
	\caption{Diagrammatic representation of the Schwinger-Dyson equation for the electron propagator. The double line represents the propagator in the nuclear
	Coulomb field, the wave line represents a virtual photon, while the thick line depicts the full, dressed electron propagator.}
	\label{fig:SDEfermion}
\end{figure}

To obtain the SDE for the bound propagator in coordinate space, we multiply Eq.~(\ref{Eq:10}) throughout with the inverse propagator, yielding
\begin{eqnarray}
	S^{-1}&&(x,y)=(i\slashed{\partial}-m-e\gamma^\mu \mathcal{A}^{0})\delta(x-y)\label{Eq:12}\\
	&&-ie^2\gamma^\mu\int \dd[4]{z}\dd[4]{w}D_{\mu\nu}(z-x)S(x,w)\Gamma^\nu(w,y;z)\,.\nonumber
\end{eqnarray}
This equation is pictorially represented in Fig.~(\ref{fig:SDEfermion}).

\section{Derivation of the self-energy shift}

The second term on the r.h.s. of Eq.~(\ref{Eq:12}), given by Eq.~(\ref{Eq:11}) gives the SE corrected
bound propagator and is the well-known SE operator. Fourier transforming this with respect to the time variable and considering the electron-photon
vertex operator $\Gamma^{\nu}$ to \textit{lowest order}, i.e. replacing it with the matrix $\gamma^{\nu}$,  we obtain
\begin{eqnarray}
	\Sigma(E)=-ie^2\int \frac{\dd{\omega}}{2\pi}\gamma^\mu S(\vb{x},\vb{w};E-\omega)D_{\mu\nu}(\vb{z}-\vb{x};\omega)\gamma^\nu\,.\label{Eq:13}
\end{eqnarray}
Using this SE operator and Feynman rules \cite{Peskin:1995ev}, we can construct the SE corrected Green's function between points $i$ and $f$:
\begin{eqnarray}
	&&G(\vb{x}_f,E_f;\vb{x}_i,E_i)\sim ie^2\bigg(\frac{i}{2\pi}\bigg)^2\int d^3\vb{z}_1 \int d^3\vb{z}_2 \int d\omega \nonumber\\
	&&\qquad\times S(\vb{x}_f,\vb{z}_2;E_f)e\gamma^\mu S(\vb{z}_2,\vb{z}_1;\eta) e\gamma^\nu\nonumber\\ 
	&&\qquad\times S(\vb{z}_1,\vb{x}_i;E_i)D_{\mu\nu}(\vb{z}_1-\vb{z}_2;\omega)\delta(E_f-E_i)\,.\label{Eq:14}
\end{eqnarray}
A Green's function, e.g. that in Eq.~(\ref{Eq:14}) can also be given in the spectral representation
\begin{eqnarray}
	G(\vb{x}_f,\vb{x}_i,E)=\sum_n\frac{\phi_n(\vb{x}_f)\bar{\phi}_n(\vb{x}_i)}{E-E_n(1-i\varepsilon)}\,\label{Eq:15}
\end{eqnarray}
in terms of the perturbed states ${\phi}_n$, where we have made the replacement $ E=E_i=E_f $, and $\varepsilon$ is infinitesimally small. The spectral function $G_a(E) \equiv \bra{a} G(E) \ket{a}$
for a given atomic reference state $\ket{a}$, which contains the perturbed eigenenergies $E_n$ of the basis states $\phi_n$, has a pole around the perturbed
eigenenergy of state $\ket{a}$:
\begin{eqnarray}
	G_a(E) \approx \frac{C_a}{E-E_a}\,,
	\label{Eq:16}
\end{eqnarray}
where the constant $C_a$ is the residue term. While in case of the path integral treatment of the quantum mechanical H atom the poles can be simply
seen in the analytical expression of the Green's function~\cite{Duru1979,PhysRevLett.53.107}, here the energies can be determined from the poles using
complex contour integration~\cite{messiah1961quantum,PhysRevA.16.863,Shabaev_2002,Shabaev1990}. Considering a small contour $\Gamma$ which surrounds an
isolated pole at the bound-state energy $E_a$, one can easily obtain
\begin{eqnarray}
	&&\frac{1}{2\pi i}\oint_{\Gamma} dE\,E\,G_a(E)= E_aC_a\,, \quad \text{and}
	\label{Eq:17}\\\nonumber\\
	&&	\frac{1}{2\pi i}\oint_{\Gamma} dE\,G_a(E)= C_a\,.
	\label{Eq:18}
\end{eqnarray}
The ratio of the above equations gives us the level energy
\begin{eqnarray}
	E_a=\frac{\frac{1}{2\pi i}\oint_{\Gamma} dE\,E\,G_a(E)}{\frac{1}{2\pi i}\oint_{\Gamma} dE\,G_a(E)}\,.
	\label{Eq:19}
\end{eqnarray}
The energy shift with respect to the unperturbed (Dirac) energy is
\begin{eqnarray}
	\Delta E_a=\frac{\frac{1}{2\pi i}\oint_{\Gamma} dE\,\Delta E\,\Delta G_a(E)}{1 + \frac{1}{2\pi i}\oint_{\Gamma} dE\,\Delta G_a(E)}\,,
	\label{Eq:20}
\end{eqnarray}
where
\begin{eqnarray}
	\Delta E_a \equiv	\Delta E_a^{(1)}+\Delta E_a^{(2)}+\dots \,,\nonumber\\
	\Delta G_a \equiv	\Delta G_a^{(1)}+\Delta G_a^{(2)}+\dots \,,\nonumber
\end{eqnarray}
have been expanded in terms of the fine-structure constant, $ \alpha $, in a perturbation series, and $ \Delta G_a= G_a(E)-\Delta G_a^{(0)} $; $ \Delta G_a^{(0)}=\frac{1}{E-E_a^{(0)}} $ being the zeroth-order contribution.
We expand the Eq.~(\ref{Eq:20}) in a geometric series and obtain
\begin{eqnarray}
	\Delta E^{(1)}_a=\frac{1}{2\pi i}\oint_{\Gamma} dE\,\Delta E\,\Delta G_a^{(1)}(E)\,.
	\label{Eq:21}
\end{eqnarray}
We can now construct the function in Eq.~(\ref{Eq:16}) from the Green's function in Eq.~(\ref{Eq:14}) and consider a single state with eigenenergy $ E_a $,
\begin{eqnarray}
	&&G_a(E)=\bigg(\frac{i}{2\pi}\bigg)^2\frac{1}{(E-E_a)^2}\label{Eq:22}\\
	&&\qquad\times\int d\omega \sum_n\frac{\bra{an}I(\omega)\ket{na}}{E-\omega-E_n(1-i\varepsilon)} \,,\nonumber
\end{eqnarray}
where we have introduced the photon exchange operator~\cite{Yerokhin1999} $ I(\vb{z}_1-\vb{z}_2;\omega)=e^2\alpha^\mu\alpha^\nu D_{\mu\nu}(\vb{z}_1-\vb{z}_2;\omega)$
with $\alpha^\mu = \gamma^0 \gamma^\mu$.
Thus, we have the SE correction to the bound-state energy level, using Eqs.~(\ref{Eq:17})-(\ref{Eq:21}), as
\begin{eqnarray}
	\Delta E_a^{(1)}=\bra{a}\Sigma(E_a)-\delta m\gamma^0\ket{a}\,,\label{Eq:23}
\end{eqnarray}
where $\bra{a}\Sigma(E_a)\ket{b}=\frac{i}{2\pi}\int d\omega \sum_n\frac{\bra{an}I(\omega)\ket{nb}}{E-\omega-E_n(1-i\varepsilon)} $; $\ket{an}$ denotes
a two-electron tensor product state
and the $ \delta m $ is the mass counter-term.

Eq.~(\ref{Eq:23}) is however fraught with divergences and we follow \cite{PhysRevA.46.3762,Yerokhin1999} and separate the SE shift into zero-, one-,
and many-potential terms
\begin{eqnarray}
	\bra{a}\Sigma(E_a)\ket{a}&=&\bra{a}\Sigma^{(0)}(E_a)\ket{a}\label{Eq:24}\\
	&+&\bra{a}\Sigma^{(1)}(E_a)\ket{a}+\bra{a}\Sigma^{(2+)}(E_a)\ket{a}\,.\nonumber
\end{eqnarray}
The individual terms in Eq.~(\ref{Eq:24}) can be written in terms of the Coulomb-Dirac Green's functions~\cite{PhysRevLett.53.107},
and the photon propagator using perturbative path integrals\cite{BHAGWAT1989417,CAMBLONG_2004,Schulman1981TechniquesAA}. The zero-potential term is given as
\begin{eqnarray}
	&&\bra{a}\Sigma^{(0)}(E_a)\ket{a}\label{Eq:25}\\
	&&=2i\alpha\int \dd{\omega}\int\dd[3]{\vb{r}_1}\dd[3]{\vb{r}_2}\psi^{\dagger}_a(\vb{r}_2)\alpha^{\mu}G^{(0)}(E_a-\omega)\nonumber\\
	&&\qquad\qquad\qquad\qquad\qquad\qquad\qquad\times\alpha^\nu D_{\mu\nu}(\omega)\psi_a(\vb{r}_1)\,,\nonumber
\end{eqnarray}
where $ G^{(0)}(E_a-\omega) $ is the free electron Green's function.

\begin{figure}
	\centering
	\includegraphics[width=0.4\textwidth]{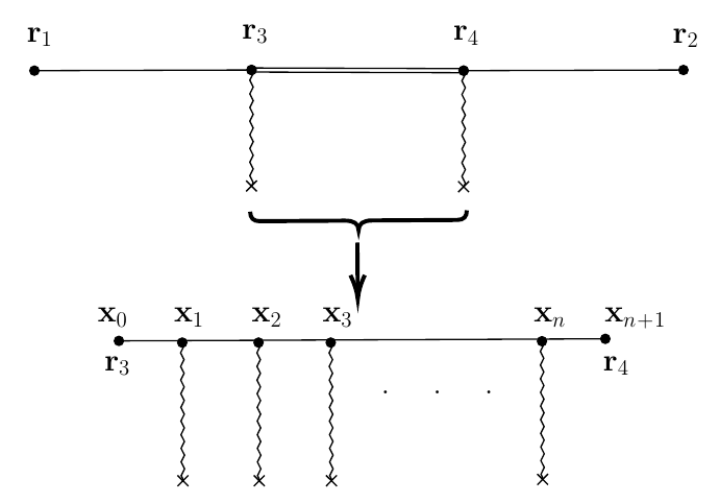}
	\caption{Diagrammatic representation of the many-potential Green's function. Single lines represent a free propagator, while wave lines terminated by a cross
	represent the nuclear Coulomb field.}
	\label{fig:GreenMP}
\end{figure}

Similarly, the one-potential term can be expressed as
\begin{eqnarray}
	&&\bra{a}\Sigma^{(1)}(E_a)\ket{a}\label{Eq:26}\\
	&&=2i\alpha\int \dd{\omega}\int\dd[3]{\vb{r}_1}\dd[3]{\vb{r}_2}\psi^{\dagger}_a(\vb{r}_2)\alpha^{\mu}G^{(1)}(E_a-\omega)\nonumber\\
	&&\qquad\qquad\qquad\qquad\qquad\qquad\qquad\times\alpha^\nu D_{\mu\nu}(\omega)\psi_a(\vb{r}_1)\,,\nonumber
\end{eqnarray}
where the Green's function for a single interaction of the electron with the nuclear potential $V$ in terms of the free Green's function
is~\cite{BHAGWAT1989417,CAMBLONG_2004,Schulman1981TechniquesAA} 
\begin{eqnarray}
	&&G^{(1)}(\vb{r}_2,\vb{r}_1;E_a-\omega)\label{Eq:27}\\
	&&=\left[\int_{0}^{\infty} \dd[3]{\vb{x_1}} V(\vb{x_1})\right]\left[\prod_{i=0}^{1}G^{(0)}(\vb{x}_{i+1}, \vb{x}_i; E_a-\omega)\right]\nonumber
\end{eqnarray}
with $\vb{r}_2 = \vb{x}_2$ and $\vb{r}_1 = \vb{x}_0$. Following the same methodology, the many-potential term is given by
\begin{eqnarray}
	&&\bra{a}\Sigma^{(2+)}(E_a)\ket{a}\label{Eq:28}\\
	&&=2i\alpha\int \dd{\omega}\int\dd[3]{\vb{r}_1}\dd[3]{\vb{r}_2}\psi^{\dagger}_a(\vb{r}_2)\alpha^{\mu}G^{(2+)}(E_a-\omega)\nonumber\\
	&&\qquad\qquad\qquad\qquad\qquad\qquad\qquad\times\alpha^\nu D_{\mu\nu}(\omega)\psi_a(\vb{r}_1)\,.\nonumber
\end{eqnarray}
The Green's function for the many-potential term between the coordinates $ \vb{r}_3 $ and $ \vb{r}_4 $, as seen in Fig.~(\ref{fig:GreenMP}),
with $ n $ insertions of the nuclear interaction where $ n\rightarrow\infty $, can be given in terms of the free Green's function as
\begin{eqnarray}
	&&G(\vb{r}_4,\vb{r}_3;E_a-\omega)\label{Eq:29}\\
	&&=\sum_{n=0}^{\infty}\Bigg\{\prod_{k=1}^{n}\left[\int_{0}^{\infty} \dd[3]{\vb{x}_k} V(\vb{x}_k)\right]\nonumber\\
	&&\qquad\qquad\qquad\qquad\times\left[\prod_{i=0}^{n}G^{(0)}(\vb{x}_{i+1}, \vb{x}_i; E_a-\omega)\right]\Bigg\}\,.\nonumber
\end{eqnarray}
This effectively gives us the exact Dirac-Coulomb Green's function~\cite{PhysRevLett.53.107}.

The energy shift due to the many-potential contribution can thus be written as
\begin{eqnarray}
	&&\bra{a}\Sigma^{(2+)}(E_a)\ket{a}\label{Eq:30}\\
	&&=2i\alpha\int \dd{\omega}\int\dd[3]{\vb{r}_1}\dd[3]{\vb{r}_2}\dd[3]{\vb{r}_3}\dd[3]{\vb{r}_4}\psi^{\dagger}_a(\vb{r}_2)\alpha^{\mu}\nonumber\\
	&&\quad\times G^{(0)}(\vb{r}_2,\vb{r}_4;E_a-\omega) V(\vb{r}_4)G(\vb{r}_4,\vb{r}_3;E_a-\omega)\nonumber\\
	&&\qquad\quad\times V(\vb{r}_3)G^{(0)}(\vb{r}_3,\vb{r}_1;E_a-\omega)\alpha^\nu D_{\mu\nu}(\omega)\psi_a(\vb{r}_1)\,.\nonumber
\end{eqnarray}
Using the spectral representation for the free and bound-electron propagators~\cite{PhysRevA.46.3762}, the energy shift is given as
\begin{eqnarray}
	&&\bra{a}\Sigma^{(2+)}(E_a)\ket{a}=\frac{i}{2\pi}\int \dd{\omega}\label{Eq:31}\\
	&&\times\sum_{\alpha,\beta, i}\frac{\bra{i} V\ket{\beta}\bra{a\beta}I(\omega)\ket{\alpha a}\bra{\alpha} V\ket{i}}{(E_a-\omega-\epsilon_{\alpha}{\varepsilon^+})(E_a-\omega-\epsilon_{i}{\varepsilon^+})(E_a-\omega-\epsilon_{\beta}{\varepsilon^+})}\,,\nonumber
\end{eqnarray}
where $ \ket{\alpha} $ and $ \ket{\beta} $ are free-electron states, the $ \ket{i} $ represents bound-electron states, and we use the notation $\varepsilon^+ = (1-i\varepsilon)$.

Following~\cite{PhysRevA.46.3762,Yerokhin1999}, we introduce frequency-dependent effective basis functions 
\begin{eqnarray}
	\ket{\phi^{(\pm)}_i(\omega)}= \sum_\alpha \frac{\bra{\alpha} V\ket{i}}{\omega-\epsilon_{\alpha}(1\mp i\varepsilon)}\ket{\alpha}\,,
	\label{Eq:32}
\end{eqnarray}
which simplify Eq.~(\ref{Eq:31}) to the form
\begin{equation}
	\bra{a}\Sigma^{(2+)}(E_a)\ket{a}=\frac{i}{2\pi}\int \dd{\omega} \sum_i\frac{\bra{a\phi_i^{(-)}}I(\omega)\ket{\phi_i^{(+)} a}}{E_a-\omega-\epsilon_{i}({1-i\varepsilon})}\,.
	\label{Eq:33}
\end{equation}
One can further reduce Eq.~(\ref{Eq:33}) by expanding the numerator inside the integral on the r.h.s. in partial waves~\cite{PhysRevA.46.3762} such that the angular
integrations can be performed analytically, yielding the reduced many-potential term expressed with generalized Slater integrals~\cite{Yerokhin1999}.
The zero- and one-potential terms with the well-known loop functions are regularized in momentum space following Ref.~\cite{Yerokhin1999}, and the numerical calculations
for the zero-, one-, and many-potential terms are performed using a B-spline representation of basis states~\cite{PhysRevA.37.307,JSapirstein_1996,Grant_2009},
implementing the dual kinetic balance approach~\cite{PhysRevLett.93.130405}.

\begin{table*}
\centering
\begin{ruledtabular}
\begin{tabularx}{\textwidth}{llllllc}
Ion & $ R_{rms} $(fm) & $1s$ & $2s$ & $2p_{1/2}$ & $2p_{3/2}$ & Ref.\\
\hline
\multirow{5}{*}{Hg$^{79+}$} & 5.475 & 206.08550(5) & 35.3186(2) &  3.155007(6) & 4.530466(6) \\
                    & 5.475 & 206.1710(1) & - & - & -  &\cite{PhysRevLett.70.158} \\
                    & 5.4648 & 206.1760(2) & 35.3919(3) & 3.23417(2) & 4.602445  &
                            \cite{2015JPCRD..44c3103Y} \\
                    & 0 & 207.18170(6) & 35.56777(1) &  3.243747(2) & 4.604480(2)  &
                            \cite{PhysRevA.46.4421} \\
                    & 0 & 207.181639 & 35.5677713(2) & 3.243742 & 4.604478  &
                            \cite{2015JPCRD..44c3103Y} \\
\hline
\multirow{4}{*}{Pb$^{81+}$} & 5.505 & 226.2388(1) & 39.2102(3)  & 3.82741(2) & 5.093(2)  \\
                    &5.5012 & 226.33237(5) & 39.28791(4) & 3.91122(2) & 5.168749  &
                            \cite{2015JPCRD..44c3103Y} \\
                    & 0 & 227.61127(5) & 39.515797(6) &  3.924620(4) & 5.171414(2) &
                            \cite{PhysRevA.46.4421}  \\
                    & 0 & 227.611278 & 39.515805 & 3.924616 & 5.171412  &
                            \cite{2015JPCRD..44c3103Y} \\
\hline
\multirow{3}{*}{U$^{91+}$} & 5.863 & 354.938(2) & 65.312(2) &  9.4392(2) & 8.853(3) \\
                   & 5.863 & - & 65.39(6) & 9.52(6) & - &\cite{PhysRevA.46.3762} \\
                   & 5.8571 & 355.052(2) & 65.4218(6) & 9.55075(6) & 8.89290(3)  &
                           \cite{2015JPCRD..44c3103Y} \\
                   & 0 & 359.50196(7) & 66.295010(9) & 9.62526(1) & 8.9010555(3) &
                           \cite{PhysRevA.46.4421} \\
                   & 0 & 359.501961 & 66.295016 & 9.625267 & 8.901058  &
                           \cite{2015JPCRD..44c3103Y}
\end{tabularx}
\end{ruledtabular}
\caption{SE correction to the energy levels of H-like heavy ions, in eV. We use a homogeneously-charged sphere as our nuclear model and compare our results
to Refs.~\cite{PhysRevA.46.3762,2015JPCRD..44c3103Y, PhysRevLett.70.158}, and point-like nucleus results from Refs.~\cite{PhysRevA.46.4421,2015JPCRD..44c3103Y}.}
\label{tab:SE}
\end{table*}

\section{Numerical results}

As a test of our method, numerical results for the SE shift in H-like HCI
are shown in Table.~\ref{tab:SE}
and compared to Refs.~\cite{PhysRevA.46.3762,PhysRevA.46.4421,2015JPCRD..44c3103Y}.
Our results are generally in good agreement with existing tabulations. We used the model of a homogeneously charged spherical nucleus, with root-mean-square
nuclear radii from Ref.~\cite{PhysRevLett.70.158}. The differences to other works originate from differences of the nuclear radii and charge distribution models used.

With the increase of the atomic number, QED effects
are boosted to the regime where they are well observable with novel mass spectrometric methods with uncertainties on the 1-eV level or
below~\cite{Kromer:2022oqq,Filianin2021,Sch_ssler_2020,Rischka2020}.
Such ions will allow, for the first time, the test of QED via measuring the mass difference of the H-like ion and the bare nucleus,
directly yielding the electronic binding energy by exploiting the energy-mass equivalence relation.
It is not only the $1s$ hydrogenic ground state which features well observable QED effects, but also excited states possess
sizeable radiative shifts. E.g. the Lamb shift of the $2s$ state approximates the SE correction to the binding energy of a Li-like ion, which
can be spectrometrically determined by measuring the mass difference of the Li- and
the He-like ions in their ground state. In our approach presented here, electron interaction effects are neglected, which, for heavy ions,
is a justified first approximation. The formalism however may be extended in future to many-electron systems, by
allowing for the exchange of photons between electrons. (We note that the Li-like sequence has been extensively studied in the framework of other
formalisms~\cite{Sapirstein2011,Kozhedub2010,Yerokhin2001}.)
Similarly, the SE results given for the $2p_{1/2}$ and $2p_{3/2}$ states (with the subscripts denoting the total angular momentum $j$) approximate the radiative
shift of the binding energy of the valence electron in the B- and N-like sequences, respectively. Very heavy ions in
these charge states still feature QED corrections accessible via mass spectrometry.

\section{Summary}

We develop an alternate formalism for the evaluation of the SE shift of atomic energy levels.
The Green's function for the SE corrected bound electron is extracted from the Schwinger-Dyson equation derived using the functional integral technique.
We avoid the operator formalism and present an elegant and intuitive framework that preserves all the symmetries of the system, and also
treats the background nuclear field non-perturbatively. This formalism can be extended to study higher-order radiative corrections,
and energy shifts in many-electron systems. The functional methods developed in this work can be naturally generalized
to various hypothetical gauge bosons, enabling the detailed study of new physics-effects in atomic spectra.

\section*{Acknowledgements}

Supported by the Deutsche Forschungsgemeinschaft (DFG, German Research Foundation) – Project-ID 273811115 – SFB 1225.

\end{document}